# Observation of Mollow Triplets with Tunable Interactions in Double Lambda Systems of Individual Hole Spins

K. G. Lagoudakis[*], K. A. Fischer, T. Sarmiento, P. L. McMahon, M. Radulaski, J. L. Zhang, Y. Kelaita, C. Dory, K. Müller & J. Vučković

*E. L. Ginzton Laboratory, Stanford University, Stanford, California 94305, United States*

Although individual spins in quantum dots have been studied extensively as qubits, their investigation under strong resonant driving in the scope of accessing Mollow physics is still an open question. Here, we have grown high quality positively charged quantum dots embedded in a planar microcavity that enable enhanced light-matter interactions. Under a strong magnetic field in the Voigt configuration, individual positively charged quantum dots provide a double-lambda level structure. Using a combination of above-band and resonant excitation, we observe the formation of Mollow triplets on all optical transitions. We find that when the strong resonant drive power is used to tune the Mollow triplet lines through each other, we observe anticrossings. We also demonstrate that the interaction that gives rise to the anticrossings can be controlled in strength by tuning the polarization of the resonant laser drive. Quantum-optical modeling of our system fully captures the experimentally observed spectra and provides insight on the complicated level structure that results from the strong driving of the double-lambda system.

The interaction of strong light fields with multi-level systems has been the subject of intense investigations over the past 40 years for both its fascinating physics and the potential for a variety of applications. Strongly-driven two-level systems are known to exhibit a fluorescence spectrum that comprises a triplet as was first predicted by Mollow [1] and demonstrated in a variety of physical systems ranging from atoms [2] and molecules [3] to superconducting qubits [4] and hybrid nanomechanical systems [5].

Another prominent system for the study of strong light-matter interactions in the solid state is the platform of semiconductor quantum dots (QDs). The atom-like behavior of quantum dots along with the simultaneous existence of excitonic and biexcitonic states, give them a very interesting cascaded Λ-V level structure that has been studied extensively under strong resonant driving [6]-[20].

Charged QDs have a spectral structure very similar to that of neutral QDs, but applying a strong magnetic field in the Voigt configuration (magnetic field perpendicular to growth axis of QD) Zeeman splits the ground and excited states, lifting the degeneracy and resulting in a double-Λ level structure.

In this Letter, we demonstrate the observation of Mollow triplets with tunable anticrossings in a double-Λ system, which originates from the resonant driving of an individual hole spin trapped in an InAs quantum dot under a strong magnetic field in the Voigt configuration.

Strong interaction of the resonant laser with the quantum dot is enabled by embedding the QDs in a planar microcavity. The microcavity was grown with Molecular Beam Epitaxy (MBE) and consists of a 25-pair AlAs/GaAs bottom Distributed Bragg Reflector (DBR), a 1-λ GaAs cavity and 5-pair AlAs/GaAs top DBR. Charging of the QDs is ensured by a beryllium (p-type) δ-doping layer, located 10 nm below the QDs. The QD layer is positioned at the center of the cavity, on an antinode of the electric field to enhance the interaction of the QDs with the resonant laser light. The resulting structure has an experimentally-estimated quality factor of Q~65 and shows strong directional emission from the asymmetric cavity. The sample structure is shown in the inset of Fig. 1(b).

The quantum dot under investigation is spectrally located close to the peak of the cavity reflection spectra, at 917.190 nm (1.351782 eV) and features a 9.8±1.7 μeV linewidth under above-band excitation at 780 nm. The photoluminescence (PL) spectra have a linear power dependence as shown by the detected photon counts using a single photon counting module (SPCM) in Fig. 1(a). Time-resolved PL measurements under weak, pulsed wetting-layer excitation at ~850 nm show a radiative lifetime of 597±2 ps yielding a Fourier-limited linewidth of 6.93 μeV, which is in good agreement with the observed PL linewidths. Applying a magnetic field in the Voigt configuration Zeeman splits the ground and excited states of the positively charged QD into a quadruplet as shown in Fig. 1(c). For this QD we find $g_h$=0.356 and $g_e$=0.416, which are typical values for this sample [21], while the diamagnetic shift factor was fitted to 3.45±0.02 μeV/T. We further investigate this quadruplet with a polarization analysis of the PL spectra at maximum magnetic field ($B_{max}$=5 T), which exhibits crossed polarizations between the inner and outer transitions as shown in Fig. 1(d). The resulting level structure is therefore a double-Λ system, as shown in the inset of Fig. 1(c). The wavy downward arrows depict the allowed transitions and their polarizations (blue-vertical and red-

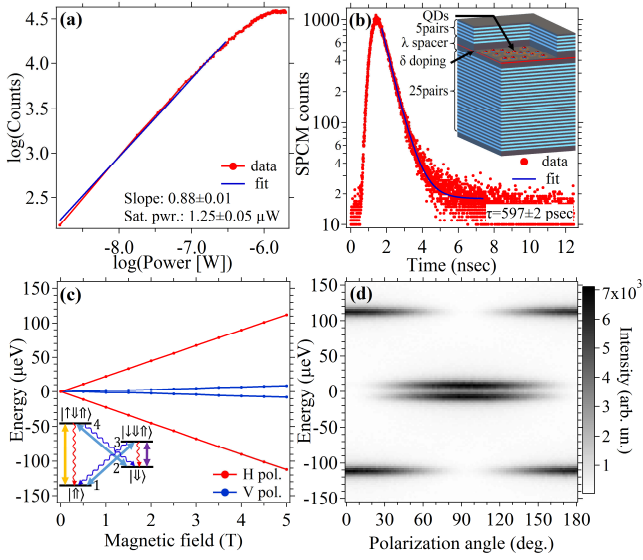

FIG. 1 (a) Above-band excitation power dependence of the emission from the investigated quantum dot. Saturation occurs at 1.25±0.05 µW driving power. (b) Time resolved QD emission under wetting layer excitation giving a lifetime of 597±2 ps. The inset depicts the sample's structure. (c) Zeeman splitting of the charged quantum dot. The inset shows the four level structure of the positively charged QD under a high magnetic field in the Voigt configuration, along with the allowed transitions and the driving schemes investigated in this work. (d) Polarization analysis of PL spectrum at 5T magnetic field.

horizontal), whereas the solid double arrows denote the excitation schemes that will be discussed later. In particular, the yellow (violet) arrow corresponds to resonant excitation of the outer highest (lowest) energy transitions (at ± 112 µeV in Fig. 1(d)) whereas the two light-blue arrows depict the excitation scheme whereby the laser is set in-between the two inner transitions (at 0 µeV in Fig. 1(d)).

In order to study the four-level system under strong resonant excitation, we pump it with a combination of two lasers: an above-band laser at saturation and a variable power resonant laser that we initially tune in-between the two inner transitions (light blue arrows in inset of Fig. 1(c) and 0 µeV in Fig. 1(d)). This excitation scheme shares similarities with a single-laser pumping and repumping scheme [22], yet here the repumping effect is almost negligible because of the above-band excitation at saturation. The above-band laser has a dual role: first it enables PL measurements, and second, it is used to stabilize the charges in the QD. When the system is driven using only resonant excitation, there is complete suppression of the positively charged exciton states. The addition of non-resonant light charges the QD and therefore enables resonant driving of the QD [23], [24]. Here, we use this feature to our advantage for the acquisition of differential spectra that provide additional suppression of the unwanted resonant laser light [21], beyond the suppression we get through the cross-polarized detection scheme. Fig. 2 shows the experimentally observed spectra as a function of the resonant driving power for four different excitation polarization angles. In particular, Fig. 2 corresponds to (a) 94° (4°), (b) 100° (10°), (c) 110° (20°) and (d) 120° (30°) excitation (detection) angles. Note that these angles are quoted with respect to the QD main polarizations, as shown in Fig. 1(d). As the detection is always cross-polarized to the excitation polarization, for the lowest detection angles, the two inner transitions and their sidebands are barely visible, while for larger angles they are easier to perceive.

The most-striking element in all four power dependences of Fig. 2 is that for increasing powers the outer transitions split into triplets −the Mollow triplets− and show linear dependence of the spectral splitting as a function of the excitation driving field up to $6 \cdot 10^{-3}$ W$^{½}$. In contrast to Mollow triplets generated by standard two-level systems, the observed triplets generated by the four-level system are centered on different energies than the laser photons. Beyond the low-power regime, the Mollow triplet splitting becomes comparable to the Zeeman splitting and an even more striking phenomenon becomes apparent: the Mollow sidebands of the inner transitions spectrally overlap with the outer transitions of the QD and instead of crossing, they show a clear anticrossing. The interaction strength between the unsplit outer transitions and the Mollow sidebands of the inner transitions can be tuned by changing the polarization of the excitation laser as shown in Figs. 2(b)-(d).

In order to understand this surprising behavior, we have to take into account the effect of the strong resonant laser on the level structure of the charged QD in the Voigt magnetic field. In the laboratory frame, the strong resonant driving leads to the dressing of both the two ground and two excited states creating an 8-level system where all transitions are allowed. A less intuitive yet more appropriate way of describing such a strongly driven system is in the rotating frame, where the system is still described as a four-level structure and its time independent Hamiltonian reads

$$H = \begin{pmatrix} -d_e/2 & 0 & \Omega\cos(\theta) & i\Omega\sin(\theta) \\ 0 & d_e/2 & i\Omega\sin(\theta) & \Omega\cos(\theta) \\ \Omega\cos(\theta) & -i\Omega\sin(\theta) & \Delta\omega - d_h/2 & 0 \\ -i\Omega\sin(\theta) & \Omega\cos(\theta) & 0 & \Delta\omega + d_h/2 \end{pmatrix} \quad (1)$$

in the $|\uparrow\Downarrow\Uparrow\rangle, |\downarrow\Downarrow\Uparrow\rangle, |\Downarrow\rangle, |\Uparrow\rangle$ basis where $d_e$ and $d_h$ are the excited and ground state Zeeman splittings, $\Omega$ is the driving field strength and $\theta$ is the laser polarization with respect to the QD main polarization axis. $\Delta\omega = \omega_o - \omega_L$ with $\omega_o$ the unsplit exciton energy for zero magnetic field, and $\omega_L$ the laser frequency.

The eigenvalues of this Hamiltonian provide a set of quasi-energies of the system, whose difference reproduces the observed PL energies along with the anticrossing behavior. The red dotted lines on top of the experimental data in Fig. 2(a)-(d)

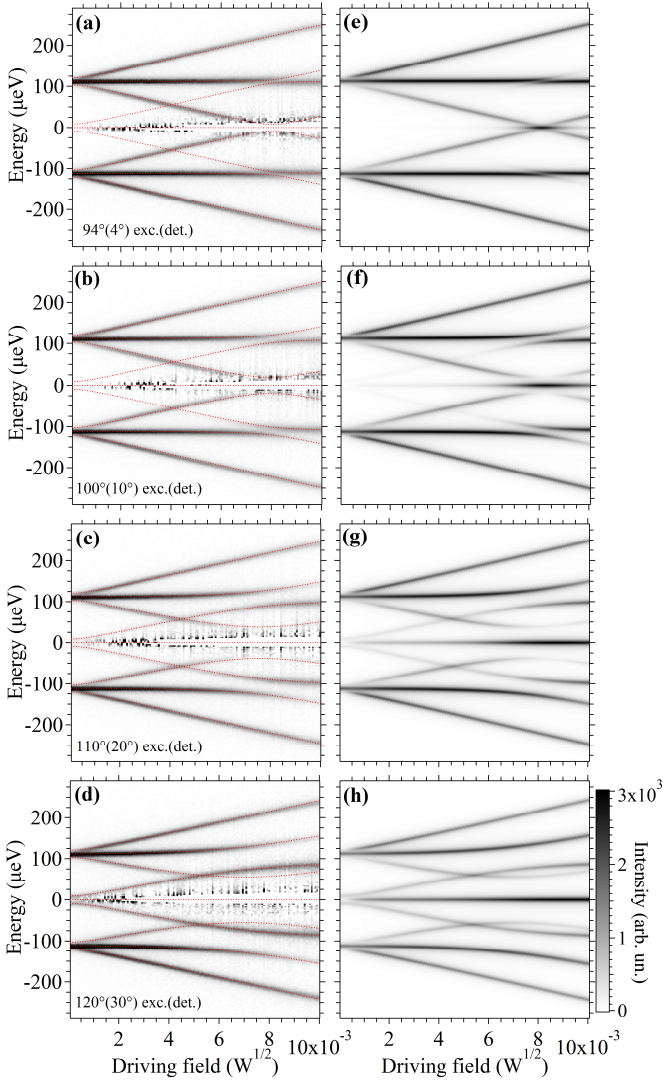

FIG. 2 (a)-(d) Power dependent experimental PL spectra for a QD strongly driven with a laser spectrally positioned in-between the two inner transitions (0 μeV), for four different excitation (detection) angles. The plots are shown in linear colorscale and have the eigenenergies of eq. (1) superposed in red dotted line. (e)-(h) Calculated spectra from quantum optical modeling for the same polarization settings as in (a)-(d) respectively.

correspond to the eigenvalues calculated from Hamiltonian in Eq. (1). Superimposing the eigenenergies on the experimental data in Figs. 2(a)-(d) enables us to further identify another interesting anticrossing, between the Mollow sidebands of the two outer transitions. The model suggests that the anticrossing splitting between the two outer-transition Mollow sidebands is twice the splitting between the inner-transition Mollow sideband and the outer-QD transition. The two aforementioned anticrossings are in stark contrast to the behavior seen for lower driving powers. At ~4.5·10$^{-3}$ W$^{½}$, where the Mollow sideband splitting is about half the Zeeman splitting, the Mollow sidebands of the inner transitions spectrally overlap with the Mollow sidebands of the outer transitions showing a clear crossing for all excitation angles.

In order to fully capture the observed phenomenology we use quantum optical modelling, here implemented using the quantum toolbox in Python (QuTiP) [25]. We solve the quantum-optical master equation of the density matrix $\rho$

$$\frac{d}{dt}\rho = -\frac{i}{\hbar}[H,\rho] + \sum_n \mathcal{L}(c_n) \qquad (2)$$

where $\mathcal{L}(c_n) = 1/2\left[2c_n\rho c_n^+ - \rho c_n^+ c_n - c_n^+ c_n \rho\right]$ is the Lindblad superoperator of the collapse operator $c_n$.

Here, $c_n = \{\sqrt{\gamma}\sigma_{13}, \sqrt{\gamma}\sigma_{14}, \sqrt{\gamma}\sigma_{23}, \sqrt{\gamma}\sigma_{24}\}$ is the spontaneous emission, while the above-band incoherent pumping is represented by $c_n = \{\sqrt{P_{ab}}\sigma_{13}^\dagger, \sqrt{P_{ab}}\sigma_{14}^\dagger, \sqrt{P_{ab}}\sigma_{23}^\dagger, \sqrt{P_{ab}}\sigma_{24}^\dagger\}$ with $\sigma_{ij}$ the two-level lowering operator, $\gamma$ the lifetime (~7 μeV), and $P_{ab} = \gamma/5$ the above-band driving power extracted from fitting.

Similar to polarizing the in-coupling by adjusting the Hamiltonian, the cross-polarized out-coupling is represented with operator $\sigma_{Xpol} = \cos(\theta)(\sigma_{13} + \sigma_{14}) + \sin(\theta)(\sigma_{23} + \sigma_{24})$ that is a weighted linear combination of all four of the double-Λ system's dipole operators. The spectra are then calculated with this new operator as the input to the standard correlator $S(\omega) = \frac{1}{2\pi}\int_{-\infty}^{\infty}\langle \sigma_{Xpol}^\dagger(t+\tau)\sigma_{Xpol}(\tau)\rangle e^{i\omega\tau}d\tau$ available in QuTiP. The model fully captures the observed phenomenology as shown in Fig. 2(e)-(h).

Additionally, we note an interesting feature of the Mollow spectra at low power that we believe is useful as a general technique for charged QD spectroscopy. For a positively-charged QD, when the outer Mollow sidebands are brighter than the inner sidebands, e.g. around 0.5 W$^{½}$, then $g_e > g_h$; this can be most clearly seen in Fig. 2(d) and (h). On the other hand, if $g_e < g_h$, then the inner sidebands would be brighter than the outer sidebands. Thus through the Mollow luminescence spectra, one can gain insight into which $g$ factor, $g_e$ or $g_h$, is the larger of the two.

To further investigate the effect of the strong laser on the four-level system of our positively charged QD, we resonantly pump the two outer transitions, as shown in Fig. 3. In Fig. 3(a) we drive the highest, whereas in Fig. 3(b) we drive the lowest energy transition. The strong laser driving leads to the dressing of the ground and excited states of the transition being driven, which in turn gives rise to a Mollow triplet, similar to that of a regular two-level system. However, this dressing also affects the two inner transitions as each of them develops into a doublet. We note that the Mollow sidebands of the driven transition show a simple crossing with the inner transition doublets.

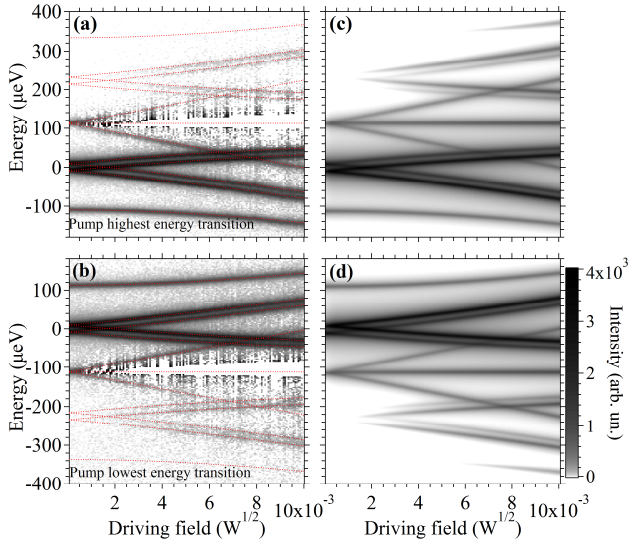

FIG. 3 (a) Power dependent spectra when the laser is in resonance with the highest energy transition of the four level system (transition 1-4 or yellow arrow in Fig. 1(c)) at 112 μeV. The colorscale is here logarithmic to highlight the "ghost" branches of the two inner transitions which appear at energies higher than any of the QD states that exist in the lab frame. (b) PL spectrum when pumping the lowest energy transition (transition 2-3 or violet arrow in Fig. 1(c)) at -112 μeV. The red dotted lines are the eigenenergies calculated from Eq. (1). (c), (d) Modeled spectra with parameters that correspond to the experimental conditions of (a) and (b) respectively.

An interesting observation is that the PL spectra also show emission at energies beyond the four transitions that exist in the laboratory frame. These "ghost" transitions are a consequence of the dressing (transitions between different dressed-state manifolds), and are fully captured by our rotating frame model as shown by the superposed eigenvalues and the calculated spectra.

In summary, we have investigated Mollow physics beyond a simple two-level system by studying the first-order correlations of the Mollow-triplet emission (the luminescence spectra) in a more intricate four-level system. We believe that our work will pave the way for future studies of nonclassical light emission that will show strong correlations in the output statistics of the sidebands. To this end, light contained in Mollow triplets has recently become of increasing interest to the community due to its ability to comprise tailored nonclassical statistics through spectral filtering [26]. We thus expect that our more complex and interacting Mollow system will provide a rich set of frequency-filtered statistics to explore in the future and might prove extremely valuable in building future nonclassical light sources.

The authors acknowledge financial support from the Army Research Office (ARO) (W911NF1310309) and the National Science Foundation (NSF) (1503759).

# Supplemental Material for

# Observation of Mollow Triplets with Tunable Interactions in Double Lambda Systems of Individual Hole Spins

K. G. Lagoudakis[*], K. A. Fischer, T. Sarmiento, P. L. McMahon, M. Radulaski, J. L. Zhang, Y. Kelaita, C. Dory, K. Müller & J. Vučković

*E. L. Ginzton Laboratory, Stanford University, Stanford, California 94305, United States*

This document provides supplemental material for the Letter "Observation of Mollow Triplets with Tunable Interactions in Double Lambda Systems of Individual Hole Spins". In particular, we provide more detailed information on the magnetic spectroscopy of the QD studied in this work as well as statistics for the charged quantum dot *g* factors. In addition we explain in detail the differential spectroscopy method that we used for the observation of Mollow triplets with tunable interactions.

**Magnetic Spectroscopy**. For the identification of the charged quantum dots we perform standard magneto-spectroscopy combined with above-band excitation. The quantum dot (QD) studied in this Letter is centered at 1.351782 eV, and features a 9.8 ± 1.7 µeV linewidth as shown by the fit in Fig. S1(a). This linewidth is very close to the resolution of our custom monochromator which is ~9 µeV. The energy difference between the two main orthogonal polarizations of this QD is fitted to 0.40 ± 0.05 µeV yielding a 0.2 µeV fine structure splitting. Application of a magnetic field perpendicular to the growth axis (Voigt configuration) splits the ground and excited states of the charged quantum dot leading to the gradual transformation of the initial doublet into a quadruplet as shown in Fig. S1(b). We used multi-peak fitting for each of the two polarizations to extract the peak locations, as is shown in Fig. S1(c). For the extraction of the linear-in-field Zeeman splitting, one has to subtract the quadratic-in-field diamagnetic shift. This is done by calculating the center of mass energy of the two outer transitions, which is shown in Fig. S1(d). A quadratic fit allows estimation of the diamagnetic shift factor, here fitted to 3.45 ± 0.02 µeV/T$^2$. Any deviation of the data from this quadratic behavior is due to thermal drifts, and as can be seen from the data in Fig. S1(d), we have taken great care to maintain a constant sample temperature (8.92 ± 0.02 K) during the adjustment of the magnetic field. To better characterize the QDs of our sample, we also performed spectroscopic studies on multiple QDs at maximum magnetic field (5T) in order to have an overview of the *g*-factors of the quantum dots. The distribution of *g*-factors is shown in Fig. S2, and is color coded according to the center wavelengths of the QD PL emission spectra.

**Differential Spectroscopy**. For the acquisition of the power dependencies that are presented in the main part of the manuscript we use differential spectroscopy as described in this section. The system under investigation is driven with two lasers; an above band laser of fixed power close to saturation (~ 1.25 µW) and a resonant laser of variable power. The above-band laser has a dual role: it is used to stabilize the charges in the QD and is used for PL emission. When the above-band laser is turned off, there is complete suppression of the charged exciton states and therefore only photons from the scattered laser are detected (due to the incomplete cross-polarized suppression). Fig. S3(a) shows the raw acquired spectra with both lasers on, while the raw spectra with only the resonant laser applied are shown in (b). A pixel by pixel subtraction of the spectra in Fig. S3(b) from those in Fig. S3(a), leaves only the PL coming from the strongly driven four level system of the positively charged QD studied in this Letter. In particular, here we show the raw data from the acquisition done with 120°/30° excitation/detection as shown in Fig. 2 (d) in the main manuscript. This method is independent of the resonant laser excitation frequency and therefore was used for the acquisition of the spectra in Figs. 2 and 3 of the manuscript.

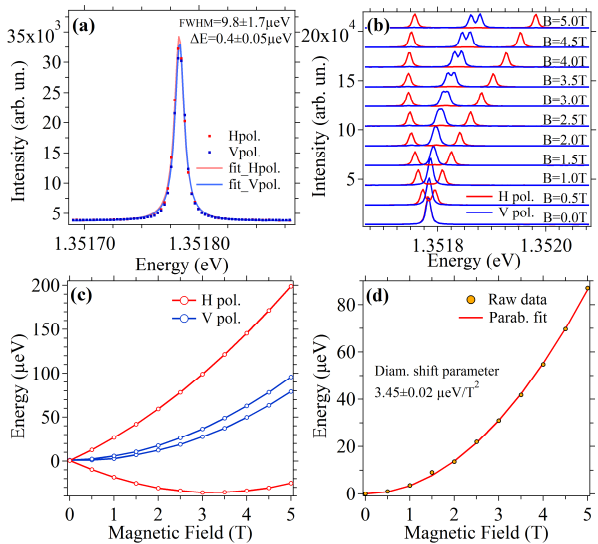

FIG. S1. (a) PL spectra of the QD studied in this Letter under zero magnetic field. (b) Waterfall plot of PL spectra for increasing magnetic field. (c) Raw peak locations for the two inner and two outer transitions extracted from multi peak fitting. (d) Center-of-mass energy of the two outer transitions. A quadratic fit allows estimation of the diamagnetic shift factor.

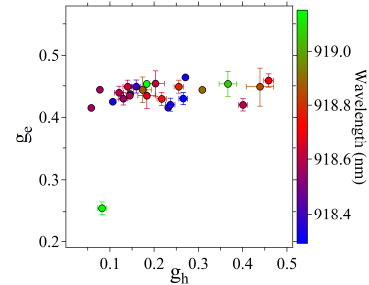

FIG. S2. Distribution of *g*-factors of ~25 quantum dots in the sample studied. The electron *g*-factors are centered around 0.43 with a narrow distribution, whereas the hole *g*-factors are distributed between 0.1 and 0.45.

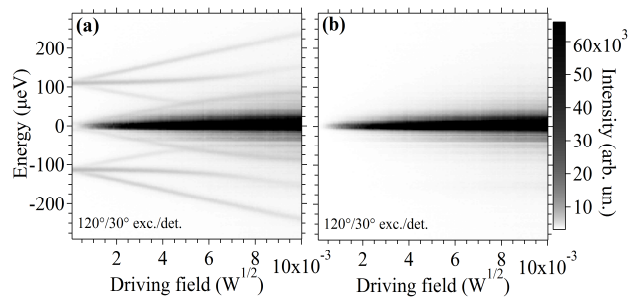

FIG. S3. (a) Raw power dependent experimental PL spectra for the QD studied in this Letter when it is driven with a combination of an above-band laser of fixed power and a resonant laser with variable power, spectrally positioned in-between the two inner transitions (0 µeV), for 120°/30° excitation/detection angles. (b) Raw power dependent spectra with only resonant laser on. Subtraction of (b) from (a) results in the data presented in the manuscript (in particular Fig. 2(d)). Note that both graphs are presented in log colorscale to enable the reader to identify the lines coming from the strongly driven four-level system.